\documentclass[prl,letterpaper,superscriptaddress,floatfix,nofootinbib,notitlepage,twocolumn]{revtex4-1}

\usepackage{textcomp}
\usepackage{amsmath}
\usepackage{amsfonts}
\usepackage{amssymb}
\usepackage{graphicx}
\usepackage{siunitx}
\usepackage{caption}
\usepackage{setspace}
\usepackage{xcolor}
\setcitestyle{super}
\usepackage[colorlinks, citecolor={blue}, linkcolor={black}]{hyperref}

\begin{document}
\raggedbottom

\title{An on-chip platform for multi-degree-of-freedom control of two-dimensional quantum and nonlinear materials}

\author{Haoning Tang}

\affiliation{School of Engineering and Applied Sciences, Harvard University, Cambridge, MA 02138, USA}

\author{Yiting Wang}
\author{Xueqi Ni}

\affiliation{School of Engineering and Applied Sciences, Harvard University, Cambridge, MA 02138, USA}

\author{Kenji Watanabe}
\author{Takashi Taniguchi}
\affiliation{National Institute for Materials Science, Namiki 1-1, Tsukuba, Ibaraki 305-0044, Japan}

\author{Pablo Jarillo-Herrero}

\affiliation{Department of Physics, Massachusetts Institute of Technology, Cambridge, MA 02139, USA}

\author{Shanhui Fan}

\affiliation{Department of Applied Physics and Ginzton Laboratory, Stanford University, Stanford, CA 94305, USA}

\author{Eric Mazur}
\email{mazur@seas.harvard.edu}
\affiliation{School of Engineering and Applied Sciences, Harvard University, Cambridge, MA 02138, USA}

\author{Amir Yacoby}
\email{yacoby@g.harvard.edu}
\affiliation{Department of Physics, Faculty of Art and Sciences, Harvard University, Cambridge, MA 02138, USA}

\author{Yuan Cao}
\email{yuancao@fas.harvard.edu}
\affiliation{Society of Fellows, Harvard University, Cambridge, MA 02138, USA}
\affiliation{Department of Physics, Faculty of Art and Sciences, Harvard University, Cambridge, MA 02138, USA}
\affiliation{Department of Electrical Engineering and Computer Science, University of California at Berkeley, Berkeley, CA 94720, USA}

\maketitle

\textbf{
Two-dimensional materials (2DM) and their derived heterostructures have electrical and optical properties that are widely tunable via several approaches, most notably electrostatic gating and interfacial engineering such as twisting. While electrostatic gating is inherently simple and has thus been ubiquitously employed on 2DM, being able to tailor the interfacial properties in a similar real-time manner represents the next leap in our ability to modulate the underlying physics and build exotic quantum devices with 2DM\cite{koren_coherent_2016,chari_resistivity_2016,ribeiro-palau_twistable_2018,yang_situ_2020,hu_-situ_2022,inbar_quantum_2023}. However, all existing approaches rely on external machinery such as scanning microscopes, which often limit their scope of applications, and there is currently no means of tuning a 2D interface that has the same accessibility and scalability as device-level electrostatic gating. Here, we demonstrate the first on-chip platform designed for 2D materials with \emph{in situ} tunable interfacial properties, utilizing a microelectromechanical system (MEMS). Each of these compact, cost-effective, and versatile devices is a standalone micromachine that allows voltage-controlled approaching, twisting, and pressurizing of two sheets of 2DM with high accuracy. As a demonstration, we engineer various flavours of synthetic topological singularities, known as half-skyrmions or merons, in the nonlinear optical susceptibility of twisted hexagonal boron nitride (h-BN)\cite{yao_enhanced_2021,yuan_synthetic_2018,ozawa_topological_2019,gobel_beyond_2021}, via simultaneous control of twist angle and interlayer separation. The chirality of the resulting moir\'e pattern further induces a strong circular dichroism in the second-harmonic generation. A potential application of this topological nonlinear susceptibility is to create integrated light sources that have widely and real-time tunable polarization, which we demonstrate experimentally. As a perspective, a quantum version of the same principle can generate entangled photon pairs with tunable entanglement characteristics. Our invention pushes the boundary of currently available technologies for manipulating low-dimensional quantum materials, which in turn opens up the gateway for designing future hybrid 2D-3D devices for applications in condensed-matter physics, quantum optics, and beyond.
}

Two-dimensional materials (2DM) emerge as a transformative class of materials with the potential to uncover novel physics and new device applications, thanks to their remarkable tunability by electrostatic gating and extensibility through van der Waals stacking\cite{novoselov_2d_2016}. The introduction of twist in stacks of 2DMs, in conjunction with the intrinsic lattice mismatch in hetereostructures, further offers control over their band structures and many-body correlations via moiré effects \cite{suarez_morell_flat_2010,bistritzer_moire_2011,lopes_dos_santos_continuum_2012,cao_correlated_2018,cao_unconventional_2018,andrei_marvels_2021,mak_semiconductor_2022}.

For nearly two decades, the assembly of 2D heterostructures has predominantly relied on dry and wet transfer methods\cite{liu_van_2016, meng_photonic_2023}. These methods have proven reliable, relatively simple to implement, and amenable to subsequent processing steps, such as etching and electrode evaporation, utilizing standard semiconductor processes. However, as the complexity of stacked 2D material structures continues to evolve, a limitation of the transfer method becomes increasingly evident. Each stacked configuration produced through transfer methods is inherently unique and non-reconfigurable\cite{lau_reproducibility_2022}. The associated non-reproducibility essentially precludes convenient exploration of various interfacial degrees of freedom (IDoFs), such as twist angle, and many conclusions can only be pulled from less than a handful of samples. 

Recent advancements in the field have witnessed the rise of scanning microscope-based platform, that can control the twist angle and perform tunneling spectroscopy at the same time\cite{inbar_quantum_2023}. While it represent a significant step forward, it is inherently limited in the scope of applicability. For example, direct optical access to the 2DMs without obstruction is not possible. Moreover, the frontier in 2DM research often requires ultra-low temperatures in combination with high magnetic fields. Reaching cryogenic temperatures and high magnetic fields in a scanning microscope is possible but challenging, as highly specialized setups are needed. These setups are inherently costly, highly complex, and often subject to vibrations from the low-temperature apparatus. Other available methods of altering the twist angle are typically not real-time with measurement and/or restricted to near-ambient conditions\cite{koren_coherent_2016,chari_resistivity_2016,ribeiro-palau_twistable_2018,yang_situ_2020,hu_-situ_2022}. 

These limitations motivate the quest for a universal approach to manipulate IDoFs in 2DM heterostructures on the device level. In response to the challenges above, we design and implement an on-chip platform based on microelectromechanical systems (MEMS) for generic manipulation of 2DM with unprecedented flexibility and accuracy. This platform, named MEms-based Generic Actuation platform for 2D materials (MEGA2D), not only addresses the pressing need for \emph{in situ} control over 2D material IDoFs but also opens a gateway to a wealth of new possibilities in condensed-matter physics, optics and beyond. MEGA2D renders the IDoFs in a 2D heterostructure as `knobs' that are now on par with electrostatic gating, in terms of convenience, scalability, and accessibility to the broader research community. The applicable area of MEGA2D can expand into almost any realm of 2DM research thanks to its versatility.

\section*{Micromachine designed for Twisting 2DM}

\begin{figure*}[!ht]
    \centering
    \includegraphics[width=\textwidth]{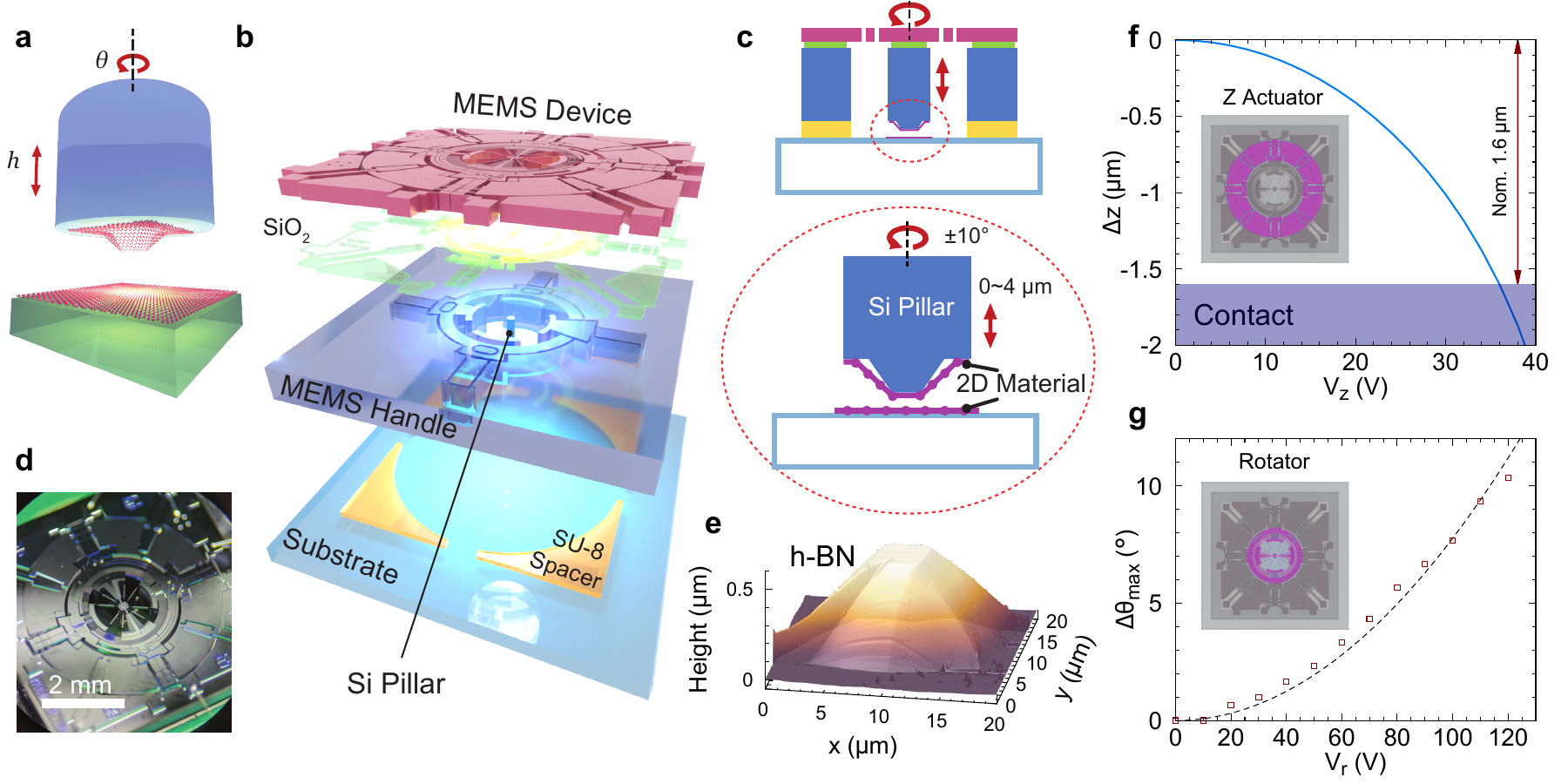}
    \caption{\textbf{MEGA2D: an on-chip MEMS platform for twisting 2D materials.} (a) 
    Illustration of the high-level idea of controlling the IDoFs in 2DM heterostructures independently using nanomechanical systems. We wish to control the twist angle $\theta$ and distance $h$ between 2DMs independently. (b) Exploded schematic of the major components in a MEGA2D device. (c) Side-cut view of an assembled MEGA2D device with layers color-coded the same as in (b). At the bottom of the central silicon pillar (circled with red dashed lines and zoomed-in in the lower panel), 2DMs are integrated, one on the pyramid etched on the bottom of Si pillar and the other on the substrate. (d) Photograph of a fully assembled MEGA2D device. (e) AFM image of an h-BN flake transferred on top of the pyramid with a height of $\sim$\SI{0.6}{\micro\meter}, exhibiting a flat surface at the top with a size of $4\times$\SI{4}{\micro\meter\squared}. (f) Calculated actuation curve of the MEMS vertical actuator. An control voltage of $V_z\sim$\SI{38}{\volt} is necessary to bring the pyramid into contact with the substrate. (g) Measured and fitted actuation curve of the MEMS rotator. A rotation up to \SI{+-10}{\degree} can be achieved at a rotational control voltage of $V_r=$\SI{120}{\volt}. The fit curve is a quadratic curve with respect to $V_r$. }
    \label{fig:fig1}
\end{figure*}

Our objective is to independently control two critical IDoFs: the twist angle, $\theta$, and the separation distance, $h$, between two layers of 2DM (Fig. 1a). Two primary physical challenges arise when utilizing MEMS for the manipulation of 2DM: (1) MEMS actuators typically exhibit limited travel ranges \cite{bell_mems_2005,algamili_review_2021}, necessitating that the 2DM layers be in close proximity, typically within a few micrometers, at their rest positions. This challenge is exacerbated when working under conditions involving low temperatures and/or high magnetic fields, where thermal expansion and magnetostriction can distort the MEMS device by more than several micrometers. To maintain platform scalability, it is imperative to preserve this initial gap without relying on an external positioner, such as a mechanical or piezo stage. (2) The mating interfaces between the 2DM layers must possess an exceptional level of flatness and parallelism to ensure consistent and uniform engagement. Ensuring this degree of parallelism over a region spanning even a few micrometres is challenging without the use of a two-axis goniometer. The introduction of such a goniometer, however, significantly increases system complexity, compromising mechanical stiffness and overall robustness.

Both of these challenges are resolved by design in MEGA2D. The schematic of a typical MEGA2D device is illustrated in Fig. 1b. Each MEGA2D device is bonded from a top silicon MEMS die and a bottom substrate using an SU-8 polymer spacer. There are two main components in each MEMS die: a \SI{60}{\micro\meter} thick silicon MEMS device layer and a \SI{450}{\micro\meter} thick silicon MEMS handle layer, bonded together by a thin layer of silicon dioxide. Figure 1c and d show a cutaway schematic and an optical image of a fully assembled MEGA2D device. At the core of the mechanism is a silicon pillar attached to the center of the MEMS device layer, which can be driven to translate vertically or rotate along its axis. At the bottom of the silicon pillar, a pyramid is etched to provide a protruding platform, where the 2DM will be transferred onto (Fig. 1c)\cite{inbar_quantum_2023}. A second piece of 2DM is placed on the substrate and laterally aligned with the pyramid, which can interact with the on-pyramid 2DM as the MEMS is vertically translated or rotated. Figure 1e shows an atomic force microscopy image of an h-BN flake on the pyramid, where the transferred 2DM forms a flat top surface, the dimension of which defined by the etched pyramid ($4\times$\SI{4}{\micro\meter\squared} in this device, but can be arbitrarily large or small down to a single point). The complete device has a dimension of about $1\times1$ \si{cm^2} and can be mounted in any room temperature or low temperature measurement apparatus, without any special equipment beyond a few electrical connections to the MEMS die. On the contrary, significant customization would be necessary in the case of scanning probe microscopes.

Since the silicon pillar and the rest of MEMS handle layer are monolithically etched from the same material (single-crystal silicon) and have exactly the same height (approximatley \SI{450}{\micro\meter}), thermal expansion and magnetostriction will have the same effect on both and their effects on the distance between the 2DMs cancel out by principle. The flatness of the 2DM interface is also guaranteed by this design, where we estimate the tilt between the 2DMs to be less than \SI{0.005}{\degree}. The free-standing distance between the 2DM is entirely determined by the difference between the thickness of the SU-8 spacer and the height of the pyramid, which equal to approximately \SI{1.6}{\micro\meter} in our experiment and highly reproducible in the same batch of devices. 

To ensure repeatable operation of MEGA2D devices, we employ electrostatic mechanisms for both vertical and rotational actuation. Unlike other types of MEMS devices that use piezoelectric and thermoelectric mechanisms, electrostatic actuators are virtually free of hysteresis and can operate at any temperatures or magnetic fields without modifications. Figure 1f shows the calculated actuation curve of the vertical actuator. Approximately \SI{38}{\volt} of control voltage is necessary to drive the pyramid into contact with the substrate. The rotational actuator has a stepper motor design\cite{stranczl_high-angular-range_2012}, which ensures frictionless and 
continuous rotation around the symmetry axis up to \SI{+-12}{\degree}. The rotation center is within \SIrange{1}{2}{\micro\meter} of the pyramid and the base step size of the stepper is $\frac{1}{3}^\circ$. Smaller angular step can be achieved by voltage interpolation but is not used in this work. Figure 1h shows the measured actuation curve for the electrostatic rotator.

We emphasize that the biggest advantage of the MEGA2D platform is its flexibility. On the MEMS die, apart from rotation and vertical displacement, other interesting IDoFs in 2DM such as lateral translation, stretching, tilting, and shearing can be incorporated as well \cite{cao_elastic_2020,zhang_topological_2020,su_topological_2020,wang_recent_2023}. Regarding the substrate, there is no limitation on the substrate material apart from its flatness. This unlocks a variety of experiments, such as optics and optoelectronics in 2DMs, which we will demonstrate below. 

\section*{Moir\'e Nonlinear Optics}

We first demonstrate the precision of the MEGA2D platform by measuring second-harmonic generation (SHG) in hexagonal boron nitride (h-BN) single-crystals. 2DMs with broken $C_2$ symmetry, such as h-BN, have strong second-order nonlinear optical response ($\chi^{(2)}$) that is highly sensitive to lattice orientation, as well as the layer arrangement along the optical path\cite{kumar_second_2013,malard_observation_2013,li_probing_2013}. Enhancement of this maximum SHG signal can be achieved via several mechanisms, including the use of cavity\cite{yi_optomechanical_2016,day_microcavity_2016,fryett_silicon_2016}, plasmonic structures\cite{akselrod_leveraging_2015}, and quasi-phase-matching\cite{yao_enhanced_2021}. 

Here we use SHG enhancement as a sensitive probe for the interlayer distance $h$ between h-BN flakes. We assemble a MEGA2D device with a h-BN flake on the pyramid and another h-BN flake on a fused silica substrate. We use a 40x objective to focus a $\lambda=$\SI{780}{\nano\meter} femtosecond laser through the fused silica substrate onto the h-BN flakes and collect the reflected SHG signal at $\lambda/2=$\SI{390}{\nano\meter}, using the same objective (Fig. 2a, see Methods for full experimental setup). The size of the pyramid top is $4\times 4$ \si{\micro\meter}, such that the light spot ($<\SI{1}{\micro\meter}$ in diameter) can be fully contained. 

We measure the SHG signal as a function of the vertical control voltage $V_z$, which tunes $h$, as shown in Fig. 2b. We find that the SHG signal is tunable by more than an order of magnitude via tuning $h$, a manifestation of the enhancement from the cavity formed by the two h-BN flakes. Remarkably, the SHG data clearly shows four periods before it settles to a constant value, which is an indication of contact between the two h-BN flakes. The cavity enhancement is periodic in second-harmonic wavelength, such that each period correspond to a motion of approximately \SI{390}{\nano\meter}, and the observed four periods therefore matches well with the designed free-standing gap of $\sim$\SI{1.6}{\micro\meter}. The reduction of the peak height as the distance increases can be attributed to decoherence of the femtosecond laser, which has a coherence length of $\sim$\SI{10}{\micro\meter}.

\begin{figure*}[!ht]
    \centering
    \includegraphics[width=\textwidth]{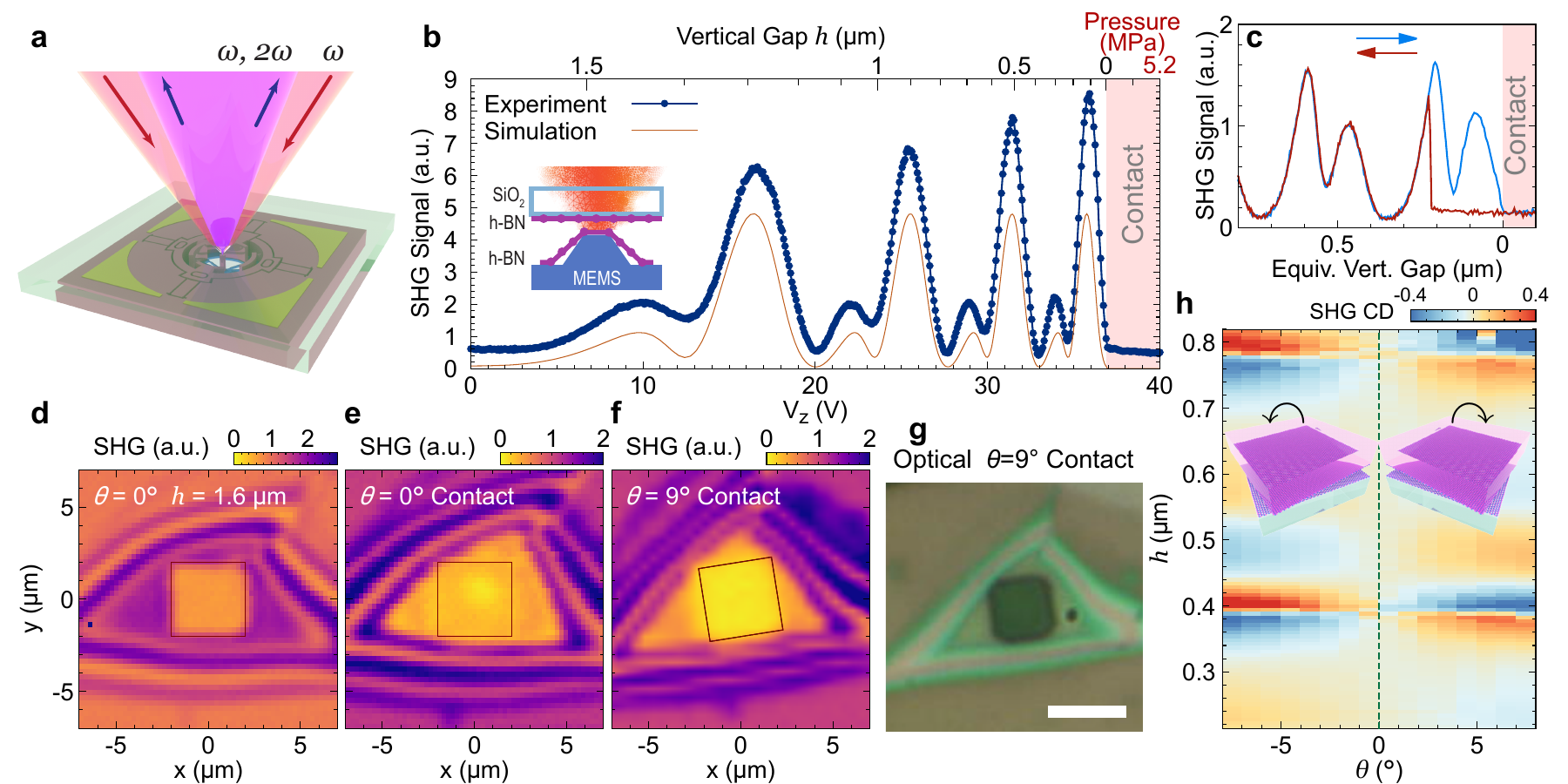}
    \caption{\textbf{Nonlinear optical probing of twisted h-BN fulfilled with MEGA2D.} (a) Schematic of the measurement scheme. A femtosecond fundamental wave at frequency $\omega$ is focused onto the backside of a MEGA2D device through its fused silica substrate, and reflected signal, part of which being at the second-harmonic frequency ($2\omega$), is collected. At the focus is two pieces of h-BN crystal, one on the fused silica and one on the MEMS pyramid tip. (b) Measured SHG power in a h-BN MEGA2D device as a function of $V_z$, the vertical control voltage of the MEMS. On the top axis, the converted separation between the two h-BN, $h$, and the applied pressure after $h$ reaches zero, is shown. The red curve shows simulated SHG power using transfer-matrix method (see Methods).
    (c) A hysteresis is seen when bringing the two h-BNs in and out of contact. The amount of hysteresis, as converted to equivalent distance, is used to deduce the van der Waals interaction force (see main text). (d) SHG power map of an as-fabricated MEGA2D device, and (e) when the h-BNs are brought into contact, which shows a possible bubble present in the middle of the pyramid top (denoted by a square). (f) Upon several rotation to $\pm\SI{10}{\degree}$ when engaged in contact, the bubble disappear and the SHG map appears highly uniform. (g) an optical micrograph of the state in (f). (h) Measured second-harmonic circular dichroism, $(P_\mathrm{SHG}^\circlearrowleft - P_\mathrm{SHG}^\circlearrowright)/(P_\mathrm{SHG}^\circlearrowleft + P_\mathrm{SHG}^\circlearrowright)$, where $P_\mathrm{SHG}^\circlearrowleft$ and $P_\mathrm{SHG}^\circlearrowright$ are the SHG powers using LCP and RCP excitations, respectively. We see that the SHG CD is antisymmetric about $\theta=0$, which is a manifestation of the inherent chirality of the moir\'e pattern.}
    \label{fig:fig2}
\end{figure*}

The built-in mechanism for maintaining the flatness of the mating interface in MEGA2D is crucial for achieving a pristine 2D interface. Figure 2d shows a typical map of SHG signal in a free-standing device ($h=\SI{1.6}{\micro\meter}$). When the vertical drive is activated and the two flakes are brought into contact (Fig. 2e), we found that the SHG signal initially became non-uniform, which can be attributed to bubbles that formed between the flakes, as typically found during stacking of 2DMs. However, upon several rotations between \SI{+-10}{\degree} while remaining in contact, the SHG signal significantly cleaned up (Fig. 2f), demonstrating the self-cleansing property of the 2D interface\cite{haigh_cross-sectional_2012,inbar_quantum_2023}.

The MEGA2D platform also allows us to directly measure mechanical properties of a 2D interface, such as the van der Waals attraction between 2DMs. These quantities are traditionally measured also by scanning probe techniques\cite{lee_measurement_2008,rokni_direct_2020}. Figure 2c shows the SHG signal measured during forward and backward scanning of the vertical control voltage as the h-BN flakes are brought into and out of contact. We find a clear hysteresis upon disengaging from contact, where the SHG signal suddenly jump from the contact value to the engagement curve, at a separation of $h=\SI{220}{\nano\meter}$. Using the calculated MEMS vertical stiffness of $k_z = $\SI{300}{\newton\per\meter}, we can estimate the maximum van der Waals attraction as $k_zd = $\SI{66}{\micro\newton}, which corresponds to a van der Waals tensile strength of \SI{4}{\mega\pascal} (contact area is assumed as \SI{16}{\micro\meter\squared}). This value is comparable but smaller than the intrinsic cleavage strength of h-BN reported in literature\cite{rokni_direct_2020}, presumably due to the misalignment between the two h-BN flakes we used in this measurement.

We now turn to the more interesting IDoF of $\theta$. h-BN crystals feature a mirror plane aligned with its armchair direction. This mirror symmetry remains intact when two h-BNs are twisted by an angle $\theta$ that is a multiple of \SI{60}{\degree}. However, for other values of $\theta$, this mirror symmetry is lifted by the emergent moiré pattern that imparts chirality to the structure. It is important to note that this alteration in symmetry is independent of the separation distance $h$; even when the h-BN crystals are not in direct contact, the chirality of the remote moiré pattern manifests observable effects.

We can directly detect this breakdown of mirror symmetry and creation of chiral response in our experiment. Chiral materials respond differently to left (LCP) and right (RCP) circularly polarized light, and its demonstration in moir\'e materials was first performed in twisted bilayer graphene\cite{kim_chiral_2016}. To probe the chirality of twisted h-BN via SHG, we insert a quarter-wave plate (QWP) before the objective to convert linear to circular polarized light (and vice versa for collected SHG wave) and compute the SHG circular dichroism (CD) defined as $(P_\mathrm{SHG}^\circlearrowleft - P_\mathrm{SHG}^\circlearrowright)/(P_\mathrm{SHG}^\circlearrowleft + P_\mathrm{SHG}^\circlearrowright)$, where $P_\mathrm{SHG}^\circlearrowleft$ and $P_\mathrm{SHG}^\circlearrowright$ are the SHG powers using LCP and RCP excitations, respectively. We purposefully fabricated a MEGA2D device with free-standing h-BN twist angle near \SI{0}{\degree}. Figure 2h shows the SHG CD, as a function of $\theta$ and $h$. We can clearly see while the measured SHG CD vary strongly with $h$, it is always antisymmetric with respect to $\theta$ and vanishes at $\theta=0$. The observed SHG CD is on the order of unity, much stronger than the CD measured in the transmittance of twisted 2DMs\cite{kim_chiral_2016}. We have thus demonstrated the ability to continuously vary the symmetry breaking properties of 2DM `on-the-fly' via MEGA2D.

\section*{Topological Texture in $\chi^{(2)}$}

The extra knobs provided by the MEGA2D platform make it possible to further study and engineer physics phenomena with new dimensions. We can treat the IDoFs offered by the MEGA2D platform as synthetic dimensions\cite{yuan_synthetic_2018,ozawa_topological_2019}. These additional continuous parameters allow higher-dimensional physics, such as topology, to be realized in a system with fewer physical dimensions. While these ideas have been recently studied extensively in photonic platforms\cite{yuan_synthetic_2018}, here we demonstrate the rich nonlinear optical properties in the synthetic space in a crystalline 2DM system, enabled by the MEGA2D technology.

In nonlinear optics involving 2DMs, the rank-3 nonlinear susceptibility tensor $\chi^{(2)}$ has eight components that are of importance. Materials with $C_3$ or higher crystallographic symmetry, such as bulk h-BN (which have additional mirror symmetry, as mentioned above) or arbitrarily twisted h-BN (no additional symmetry), have two independent groups of coefficients: $\chi_1 = \chi_{xxx}^{(2)}=-\chi_{yyx}^{(2)}=-\chi_{yxy}^{(2)}=-\chi_{xyy}^{(2)}$ and $\chi_2 = -\chi_{yyy}^{(2)}=\chi_{xyx}^{(2)}=\chi_{xxy}^{(2)}=\chi_{yxx}^{(2)}$\cite{boyd_nonlinear_2020}. We can write these two components in a compact pseudospin form: $\psi = [\chi_1, \chi_2]$. For bulk h-BN with 
mirror symmetry ($D_{3h}$ symmetry), $\psi$ is proportional to $[1,0]$ if $x$-direction is parallel to its armchair direction. 

\begin{figure*}[!ht]
    \centering
    \includegraphics[width=\textwidth]{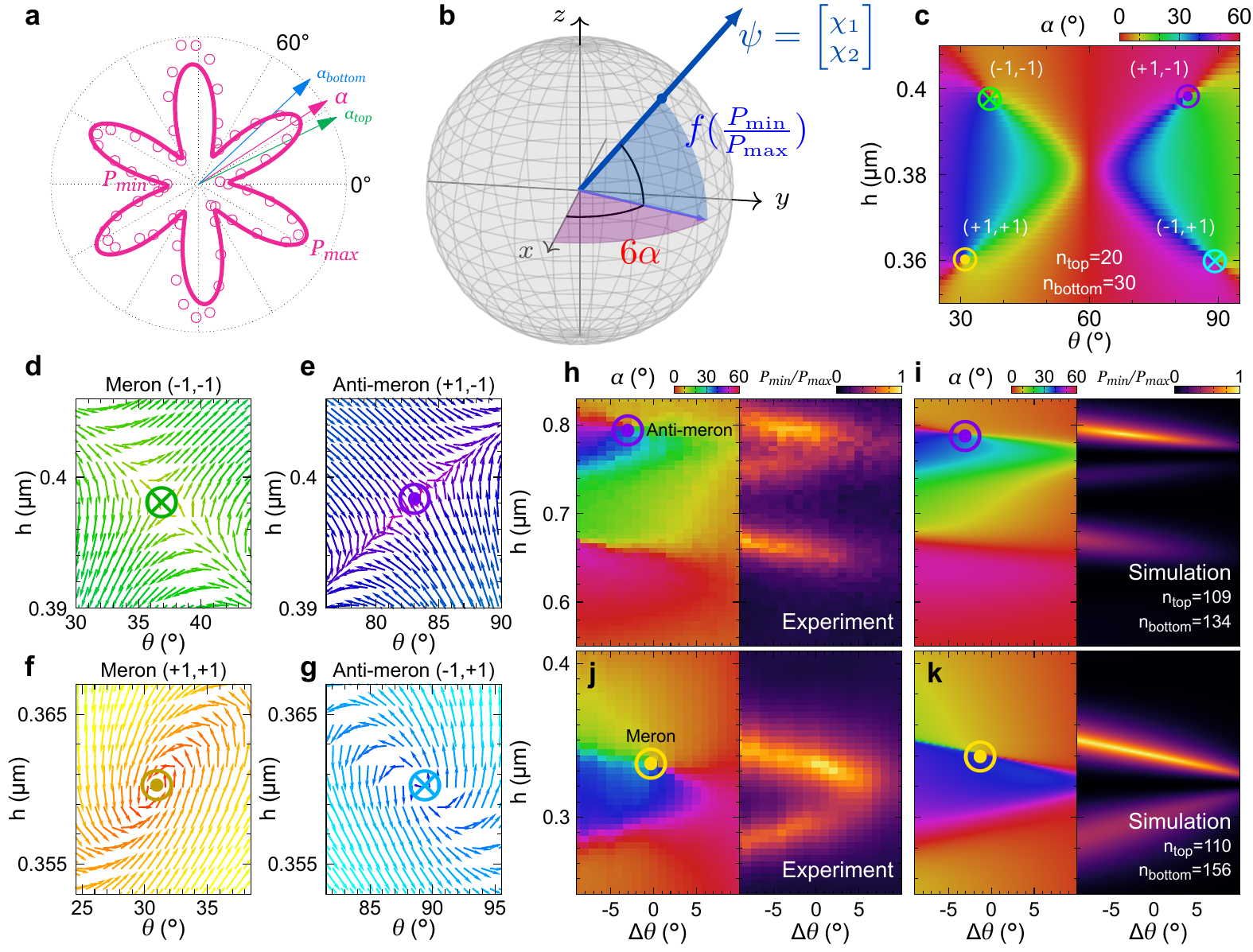}
    \caption{\textbf{Experimental realization of synthetic merons (half-skyrmions) in the nonlinear susceptibility of twisted h-BN.} (a) A typical polarization-dependent SHG measurement in twisted h-BN. From each curve like this, we can fit the SHG power to a cosine function with respect to the polarization to obtain three independent parameters: SHG power maxima $P_\mathrm{max}$, minima  $P_\mathrm{min}$, and polarization $\alpha$ of the maxima. We also showed the polarization of maxima $\alpha_\mathrm{top}$ and $\alpha_\mathrm{bottom}$, if only top and bottom h-BN were to be measured. (b) The effective nonlinear susceptibility of twisted h-BN, if written as a nonlinear pseudospin $\psi$, represents a direction on the Poincar\'e sphere. The azimuth and elevation of this direction can be extracted from measured $\alpha$ and $P_\mathrm{min}/P_\mathrm{max}$, respectively. $x,y,z$ directions represent nonlinear pseudospins $[1,0], \frac{1}{\sqrt{2}}[1,1], \frac{1}{\sqrt{2}}[1,i]$ respectively. The function $f(x)=2\tan^{-1}(x^{1/2})$. (c) shows the numerical evolution of $\alpha$ in the synthetic space for a system with $n_\mathrm{top}=20$ and $n_\mathrm{bottom}=30$. In this space, we found four types of merons, labeled as $(p,v)$ by their core polarity $p$ and vorticity $v$. $p=+1$ (-1) means at the center of the meron or anti-meron, $\psi$ points towards $+z$ ($-z$), which is denoted by symbols with solid dot (cross) at the center. $v=+1$ (-1) means that $\alpha$ increases counterclockwise (clockwise). The pseudospin texture of the four types of merons are shown in (d-g) in more detail. The product of $p$ and $v$ defines the topological charge $Q=\frac{1}{2}pv$, which equals $+\frac{1}{2}$ for a meron and $-\frac{1}{2}$ for an anti-meron. (h) shows the experimentally measured $\alpha$ and $P_\mathrm{min}/P_\mathrm{max}$ and (i) shows the numerical simulation for a MEGA2D device, which exhibits a $(+1,-1)$ anti-meron in the accessible synthetic space. The device has an initial twist of $\theta=\SI{43}{\degree}$ at $\Delta\theta=0$. (j-k) shows the experiment and simulation for a similar device that possesses a $(+1,+1)$ meron instead.}
    \label{fig:fig3}
\end{figure*}

For a generic stack of h-BN (or any 2DM with parent three-fold symmetry), the effective nonlinear susceptibility is a phase-coherent superposition of the susceptibility of each constituent layer, and the effective $\psi$ represents a direction on the Bloch (Poincar\'e) sphere (see Fig. 3b). We can view this direction as the orientation of the pseudospin vector. In twisted h-BN, this vector is sensitively dependent on $h$ and $\theta$, as well as the number of layers in each h-BN flake. 
In particular, $\psi$ is periodic in both $h$ (periodicity is $\lambda/2$, as we eluded in Fig. 2b) and $\theta$ (periodicity is \SI{120}{\degree}), and the $h$-$\theta$ synthetic space has the same torus-like topology as the first Brillouin zone in a 2D crystal lattice. As we show in the Methods, a direct physical manifestation of the pseudospin orientation is that it defines a polarization state (generally elliptical), at which maximal nonlinear interaction is achieved, \emph{e.g} a maximum in SHG power. Essentially, the second-order nonlinear optics in twisted $C_3$-symmetric 2DMs realizes an effective two-band system in the $h$-$\theta$ synthetic space. The spectra and eigenstates of this two-band system represents the SHG power and its polarization state, respectively. 

Similar to spin textures found in various magnetic materials\cite{gobel_beyond_2021,guo_meron_2020}, the nonlinear pseudospin can also exhibit nontrivial textures in the synthetic space. Figure 3c shows the calculated values of $\alpha$ in the synthetic space, for $n_\mathrm{top}=20$ layers of h-BN on the fused silica and $n_\mathrm{bottom}=30$ layers of h-BN on the silicon pyramid respectively. Rich features exist at certain locations of the synthetic space, where $\alpha$ acquires a rotation from \SI{0}{\degree} to \SI{60}{\degree} (or vice versa) around singularities. These singularities are analogous to topological quasiparticles known as half-skyrmions or merons\cite{gobel_beyond_2021,guo_meron_2020}. The merons are characterized by their topological charge $Q=\frac{1}{2}p\cdot v=\pm1/2$, which is determined by the core polarity $p=\pm1$ and the vorticity $v=\pm1$. $p=+1$ or $-1$ means that $\psi$ points to $+z$ or $-z$ on the Poincar\'e sphere at the core of the meron, whereas $v=+1$ or $-1$ means that $\psi$ forms a vortex or anti-vortex around that core. There are therefore four possible types of merons/anti-merons in total. In our system, we theoretically found all four types of synthetic merons ($Q=+1/2$) and anti-merons ($Q=-1/2$), as illustrated in Fig. 3d-g. 

With the tunable MEGA2D platform, we can now access these rich topological features experimentally. To probe the nonlinear pseudospin $\psi$, we measure the SHG power as a function of linear polarization in a parallel configuration, where incident and reflected SHG waves have the same polarization. This type of measurements yields the characteristic six-fold symmetric pattern with respect to the incident polarization\cite{kumar_second_2013,malard_observation_2013,li_probing_2013}, as shown in Fig. 3a. We can identify three quantities from each measurement: the maxima $P_\mathrm{max}$, the minima $P_\mathrm{min}$, and the polarization angle of the maxima $\alpha$, which is in the range of $\SI{0}{\degree}\leq\alpha < \SI{60}{\degree}$. These quantities are directly tied to the geometrical representation of $\psi$ on the Poincar\'e sphere (see Fig. 3b). The azimuthal angle of $\psi$ on the sphere equals $6\alpha$, and its altitude is given by $2\tan^{-1}(P_\mathrm{min}/P_\mathrm{max})^{1/2}$. 

Figure 3h-k show the measured and calculated polarization angle $\alpha$ and SHG power ratio $P_\mathrm{min}/P_\mathrm{max}$ in two MEGA2D devices. In both devices the experiments can be well-modeled by our simulation. In one device (Fig. 3h-i), we can clearly identify an anti-meron ($Q=-1/2$) labeled by $(p,v)=(+1,-1)$. At the core of the anti-meron, we find that $P_\mathrm{min}/P_\mathrm{max}$ approaches unity, consistent with the expectation that $\psi$ point towards poles of the Poincar\'e sphere. In another device with different numbers of h-BN (Fig. 3j-k), we found a meron of the type $(+1,+1)$ with similar behaviours. We note that the polarity $p$ of the meron is not directly determinable by linear polarization SHG, but can be separately identified via SHG CD measurement (see Methods).

The pseudospin texture of an isolated meron covers half of the Poincar\'e sphere, and that of a meron-meron pair, which is topologically equivalent to a full skyrmion, in principle have a full coverage on the Poincar\'e sphere. A practical implication of the existence of these synthetic topological quasiparticles is that we can now engineer the second-order susceptibility $\chi^{(2)}$ of an active optical stack to cover the full available space permitted by symmetry, while $h$ and $\theta$ only needs to be tuned in a small range. We will next show how this could be exploited to explore new applications of nonlinear optics using MEGA2D.

\section*{Towards a tunable quantum light source}

The ultimate tunability of the nonlinear susceptibility near a synthetic meron could be utilized to design full-Stokes tunable classical and quantum light sources. 

A second-order nonlinear optical interaction involves three photons with energies $\hbar\omega_1, \hbar\omega_2$, and  $\hbar\omega_3$, where $\hbar$ is the reduced Planck constant and $\omega_1+\omega_2=\omega_3$. These photons also define three polarization states, $\alpha_1$, $\alpha_2$, $\alpha_3$. In a bulk nonlinear material, these polarizations are typically fixed to certain directions that are parallel to its crystallographic planes in order to satisfy phase-matching requirements\cite{boyd_nonlinear_2020}. In nonlinear 2DMs, however, the phase-matching requirement is relaxed and the polarization states of these photons are connected by a fourth polarization defined by the nonlinear pseudospin, $\gamma=\tan^{-1}(\chi_2/\chi_1)$, according to the simple sum rule \cite{boyd_nonlinear_2020,trovatello_optical_2021}
\begin{equation}
    \label{eq:sumrule}\alpha_1 + \alpha_2 + \alpha_3 = \gamma + k\pi,\quad k=0,\pm1,\pm2,\ldots.
\end{equation}
Here, different polarization states are represented by complex numbers with the real parts denoting their polarization angle and imaginary parts denoting their ellipticity, as mapped by a Mercator conformal projection from the Poincaré sphere (see Methods).

\begin{figure*}[!ht]
    \centering
    \includegraphics[width=\textwidth]{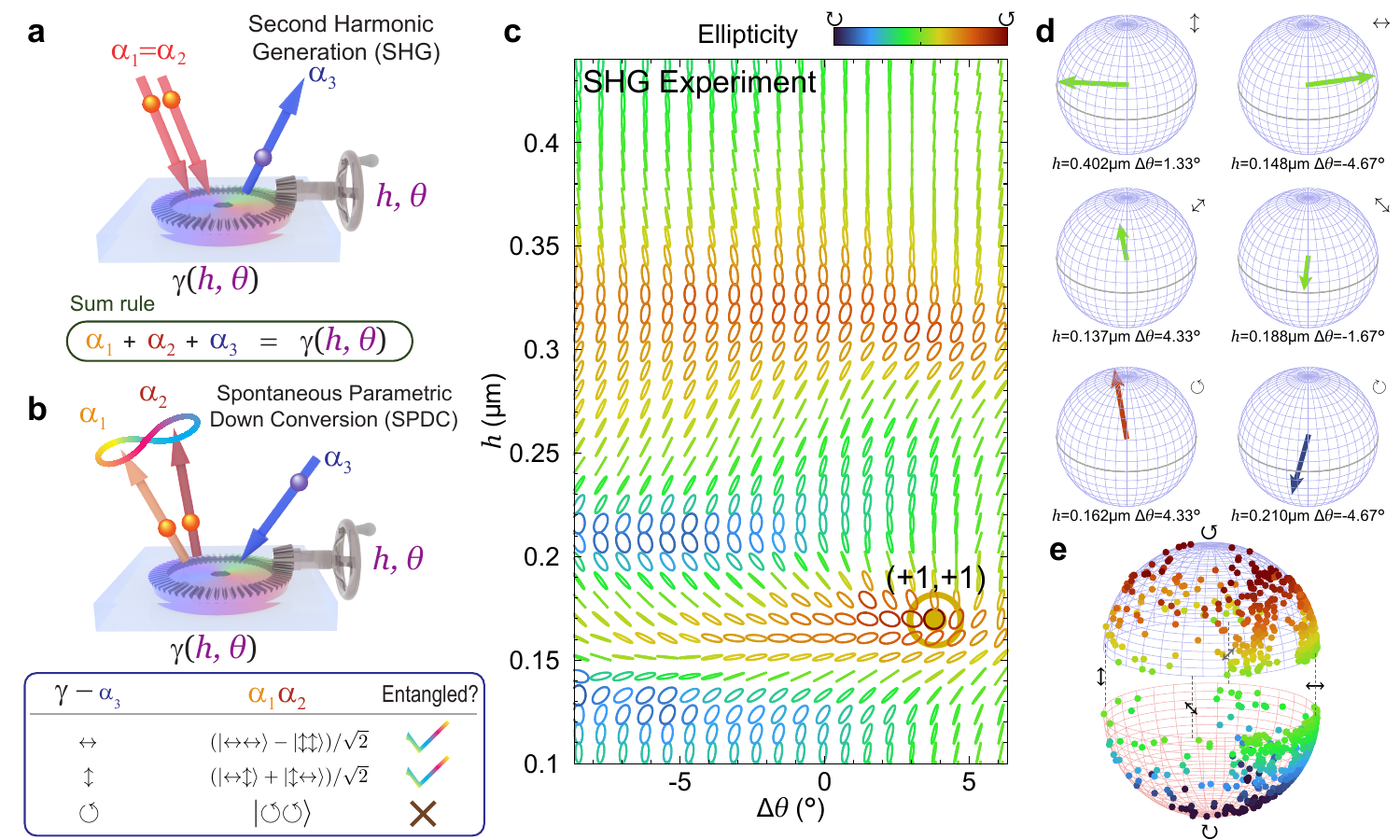}
    \caption{\textbf{Tunable classical and quantum light source with MEGA2D.} Twisted h-BN has an intrinsic dimensionless complex number $\gamma=\tan^{-1}(\chi_2/\chi_1)$ that characterize its nonlinear susceptibility. (a) In SHG process, two photons with polarization $\alpha_1$ and $\alpha_2$ (typically equals $\alpha_1$) combine into a photon with polarization $\alpha_3$. These angles obey a sum rule, so that $\alpha_3$ can be controlled by $\gamma$, which is in turn tunable by $h$ and $\theta$. (b) In spontanoues parametric down-conversion (SPDC), which is the inverse process of SHG, the same sum rule is obeyed. However, $\alpha_1$ can generally be different from $\alpha_2$, which results in an ambiguity. This ambiguity is related to the inherent quantum randomness of the SPDC process, and what we obtain is entangled photon pairs as shown in the table. The strength and direction of the entanglement is tunable by $\gamma$ and by extension $h$ and $\theta$. Experimentally we demonstrate only the SHG process here. (c) Demonstration of tunable SHG light source with MEGA2D. We show the measured $\alpha_3$ as a function of $h$ and $\theta$. Each ellipse in this plot represents a polarization state, with colour representing its ellipticity. The $(+1,+1)$ synthetic meron located near the lower right corner of this diagram allows us to tune $\alpha_3$ into left circular polarization. (d) shows that the six major polarizations that we can access in this MEGA2D device, and (e) shows the coverage of the Poincar\'e sphere. The upper and lower halves of the Poincar\'e sphere is split for clarity. The lower hemisphere shows less coverage because the accessible meron has $p=+1$.}
    \label{fig:fig4}
\end{figure*}

With $\gamma$ being tunable via the IDoFs $h$ and $\theta$, we can realize a light source at $2\omega$ frequency with tunable polarization state $\alpha_3$. In bulk nonlinear crystals, the polarization of SHG cannot be conveniently tuned due to phase-matching conditions. In previous experiments on 2DMs, active tuning of SHG polarization can only be achieved via intricate control of the arrival times of the two fundamental photons, and the tuning range is limited to a subset of polarizations\cite{klimmer_all-optical_2021}. The tunability of $\psi$ in our case can in principle span the entire Poincaré sphere, and a full-Stokes tunable SHG source can thus be implemented, with $h$ and $\theta$ being the only necessary knobs. We fix the fundamental laser beam to vertical polarization $\alpha_1=\alpha_2=0$ and measure the SHG polarization $\alpha_3$ using quarter-wave plate polarimetry (see Methods) in a MEGA2D device, as shown in Fig. 4c. We can directly see the polarization vortex in $\alpha_3$ at a synthetic meron, around which $\alpha_3$ rotates by \SI{180}{\degree}. Exactly at the center of the vortex, the SHG wave becomes left circularly polarized; this can be described by $\alpha_3=\gamma = +i\infty$ in Eq. \ref{eq:sumrule}. By tuning $h$ and $\theta$, we could achieve nearly any polarization state; Fig. 4d shows that all six principle polarizations on the Poincar\'e sphere can be approximately reached within the accessible synthetic space, and Fig. 4e shows its coverage on the sphere. Future generations of MEGA2D devices that allow access to a synthetic full skyrmion could provide an even better coverage of the Poincar\'e sphere. 

As a future perspective, another exciting possibility using the same principles and MEGA2D platform lies in the reciprocal process of SHG, which is known as spontaneous parametric down-conversion (SPDC), that converts a single photon at $2\omega$ to two photons at $\omega$\cite{zhang_spontaneous_2021}. While SHG is fundamentally classical, SPDC is an inherently quantum process and can create entangled photon pairs with high purity. SPDC process relies on the same $\chi^{(2)}$ tensor, and therefore the same sum rule Eq. \ref{eq:sumrule} applies. If we fix incident $\alpha_3$ and $\gamma$, however, $\alpha_1$ and $\alpha_2$ can take arbitrary values as long as their sum obeys Eq. \ref{eq:sumrule}. The result is entangled photons, the entanglement of which is further tunable by varying $\gamma(h,\theta)$ (see Fig. 4b). If $\alpha_1+\alpha_2=0$, we obtain a Bell's state $(\left|\leftrightarrow\leftrightarrow\right> - \left|\updownarrow\updownarrow\right>)/\sqrt{2}$ (see Methods for rigorous derivation). If $\alpha_1+\alpha_2=\pi/2$, we obtain another Bell's state $(\left|\leftrightarrow\updownarrow\right> + \left|\updownarrow\leftrightarrow\right>)/\sqrt{2}$. If $\gamma=+i\infty$ (i.e. $\psi\propto [1,i]$), however, $\alpha_1$ and $\alpha_2$ are both forced to left circular polarization and the resulting photons have zero entanglement ($\left|\circlearrowleft\circlearrowleft\right>$ state). This flexibility of tuning the polarization and the extent of entanglement is, to our knowledge, not available in other SPDC platforms, because it takes advantage of the fact that SHG/SPDC in 2DMs does not rely on double refraction and are not restricted by phase-matching requirements. Such a device, when implemented with MEGA2D, could enable further compactification of quantum optics instruments. We believe that a 2DM with a stronger nonlinear response than h-BN would be necessary to experimentally realize such a tunable quantum light source, possible candidates include 3-R transitional metal dichalcogenides and other non-centrosymmetric 2DMs\cite{shi_3r_2017,guo_ultrathin_2023}.

As we showed in several demonstrations, MEGA2D represents a transformation in the methodology of studying 2DMs and enables a plethora of new directions to be explored when the IDoFs become freely tunable, even with a previously well-characterized material. We envision that the unlimited combination of existing and new 2DMs with MEGA2D will reveal numerous new physics as well as device applications in the near future. The same technique can also be seamlessly integrated with flat metamaterials such as photonic crystals and unlock new ways to manipulate light and study light-matter interactions\cite{tang_experimental_2023}.

\section*{Acknowledgements}

We thank for helpful discussions with Mingjie Zhang, Beicheng Lou, and Guixiong Zhong. H. T. and E. M. acknowledges support from DARPA under contract URFAO: GR510802. A.Y. acknowledges support from the Army Research Office under Grant number W911NF-21-2-0147 and the Gordon and Betty Moore Foundation through Grant GBMF 9468. S.F. acknowledges the support of a MURI grant from the U.S. Air Force Office of Scientific Research (grant no. FA9550-21-1-0312). The sample fabrication was performed at Harvard University Center for Nanoscale Systems, which is a member of the National Nanotechnology Coordinated Infrastructure Network and is supported by the National Science Foundation under NSF award 1541959. K.W. and T.T. acknowledge support from the Elemental Strategy Initiative conducted by the MEXT, Japan, grant no. JPMXP0112101001; JSPS KAKENHI grant no. JP20H00354; and the CREST (JPMJCR15F3), JST. PJH acknowledges support by the National Science Foundation (DMR-1809802), the STC Center for Integrated Quantum Materials (NSF Grant No. DMR-1231319), the Gordon and Betty Moore Foundation’s EPiQS Initiative 
through Grant GBMF9463, the Fundación Ramon Areces and the CIFAR Quantum Materials program.

\section*{Author Contributions}
Y. C. and A. Y. conceived the experimental ideas. Y. C. designed the MEGA2D platform. H. T., Y. W. and Y. C. performed the sample fabrication. H. T., E. M. and Y. C. performed the optical measurements. K. W. and T. T. provided h-BN crystals. X. N., S. F. and Y. C. developed the theory. S. F and P. J. H. provided insightful discussions. H. T. and Y. C. wrote the manuscript with input from all authors.

\section*{Author Information}
The authors declare no competing financial interest.

\end{document}